\documentstyle[preprint,psfig,epsfig,12pt]{aastex}

\def\gs{\mathrel{\raise0.35ex\hbox{$\scriptstyle >$}\kern-0.6em
\lower0.40ex\hbox{{$\scriptstyle \sim$}}}}
\def\ls{\mathrel{\raise0.35ex\hbox{$\scriptstyle <$}\kern-0.6em
\lower0.40ex\hbox{{$\scriptstyle \sim$}}}}

\lefthead{A.C.\ Edge et al.}  \righthead{A\,1201 - A unique arc}

\begin{document}

\title{A unique small--scale gravitational arc in Abell~1201$^1$}

\author{ Alastair C.\ Edge,$\!$\altaffilmark{2} Graham P.\
  Smith,$\!$\altaffilmark{3} David J.\ Sand,$\!$\altaffilmark{3}
  Tommaso Treu,$\!$\altaffilmark{3,4,5} Harald
  Ebeling,$\!$\altaffilmark{6} Steven W.\ Allen,$\!$\altaffilmark{7}
  Pieter G.\ van~Dokkum,$\!$\altaffilmark{3,8}}

\altaffiltext{1}{Based on observations at the Keck Observatory, which is
      operated jointly by the California Institute of Technology and
      the University of California, and with the NASA/ESA Hubble Space
      Telescope, obtained at Space Telescope Science Institute, which
      is operated by the Association of Universities for Research in
      Astronomy, under NASA contract NAS5-26555.}
\altaffiltext{2}{Department of Physics, University of Durham, South
      Road, Durham DH1 3LE, UK}  
\altaffiltext{3}{California Institute of Technology, Department of
      Astronomy, Mail Code 105--24, Pasadena, CA91125, USA} 
\altaffiltext{4}{Department of Physics and Astronomy, University of 
California at Los Angeles, Los Angeles, CA 90095, USA}
\altaffiltext{5}{Hubble Fellow}
\altaffiltext{6}{Institute for Astronomy, 2680 Woodlawn Drive,
      Honolulu, HI 96822, USA} 
\altaffiltext{7}{Institute of Astronomy, University of Cambridge,
      Madingley Road, Cambridge CB3 0HA, UK}
\altaffiltext{8}{Department of Astronomy, Yale University, New Haven,
      CT 06520, USA} 
\setcounter{footnote}{6}

\begin{abstract}
We present a snapshot \emph{Hubble Space Telescope (HST)} image of the
galaxy cluster A\,1201 ($z{=}0.169$), revealing a tangential arc
2\,arcsec from the brightest cluster galaxy (BCG).  Keck--ESI
spectroscopy confirms that the arc is gravitational in nature and that
the source galaxy lies at $z{=}0.451$.  We construct a model of the
gravitational potential of the cluster that faithfully reproduces the
observed arc morphology.  Despite the relaxed appearance of the
cluster in the \emph{HST} frame, the best fit ellipticity of the total
matter distribution is $\epsilon_{\rm total}{\ge}0.5$, in contrast to
the light distribution of the BCG ($\epsilon_{\rm
BCG}{=}0.23{\pm}0.03$) on $2''$ scales.  Further deep optical
observations and pointed X--ray spectro--imaging observations with
\emph{Chandra} are required to determine whether this elongation is
due to a single elongated dark matter halo, or a more complex
distribution of matter in the cluster core.  We compare the arc with a
sample drawn from the published literature, and confirm that it is
unique among tangential systems in the small physical scales that it
probes (${\sim}6$\,kpc).  In anticipation of a more thorough
investigation of this cluster across a broad range of physical scales,
we use our fiducial lens model to estimate the projected mass and
mass--to--light ratio of the cluster within a radius of 6\,kpc,
obtaining: $M{=}(5.9^{{+}0.9}_{{-}0.7}){\times}10^{11}{\rm M_\odot}$,
$M/L_V{=}9.4^{{+}2.4}_{{-}2.1}(M/L)_\odot$.  Overall our results confirm the
importance of \emph{HST} snapshot surveys for identifying rare lensing
constraints on cluster mass distributions.  In combination with
follow--up optical and X--ray observations, the arc in A\,1201 should
help to increase our understanding of the physics of cluster cores.
\end{abstract}


\section{Introduction}

Galaxy clusters are important laboratories in which to study physical
processes that are generally inaccessible in other environments.  For
example the radial density profile and the projected ellipticity of
clusters on the sky may offer valuable clues into the nature of dark
matter (e.g.,\ Spergel \& Steinhardt 2001; Sand et al.\ 2002; 2003;
Miralda-Escud\'e 2002; Arabadjis et al.\ 2002).  Complications often
arise in cluster--based studies of dark matter due to the presence of
baryons (e.g.,\ Allen 1998; Smith et al.\ 2001; Lewis et al.\ 2003).
However, from a broader perspective such complications provide us with
important clues into the physics of gas cooling, and inter--play
between baryons and dark matter, both of which are central to
attempts to understand the physics of galaxy formation (e.g.,\ Cole
et al.\ 2000).

Progress towards these goals requires detailed study of the
distribution of mass in clusters.  Strong gravitational lensing offers
a direct and precise probe of cluster mass distributions (e.g.,\ Kneib
et al.\ 1996; Smith 2002; Smith et al.\ 2003).  Complementary
constraints can also be obtained from X--ray observations (e.g.,\ Allen
et al.\ 2002), weak lensing (e.g.,\ Kneib et al.\ 2003) and the
three--dimensional distribution of cluster galaxies (e.g.,\ Czoske et
al.\ 2002).  A combination of these techniques is necessary for a
comprehensive understanding of mass in clusters.  Armed with the
results from such multi--wavelength studies, robust constraints on the
dark matter particle and gas cooling may ultimately flow.

We have conducted a snapshot survey of 55 X--ray luminous galaxy
clusters with the WFPC2 camera on--board \emph{HST} (PID's 8301 \&
8719; PI Edge).  A key goal of this survey is to uncover new cluster
lenses with which to explore the questions outlined above.  The
snapshot observing strategy is well--suited to 
identifying clusters containing rare and powerful
constraints such as radial arcs (e.g.,\ RXJ\,1133 -- Sand et al.\ 2003).

%\begin{sloppypar}
In this letter we present an \emph{HST} observation of A\,1201
($z{=}0.169$;
$(\alpha,\delta){=}11^h13^m01.1^s{+}13^\circ25'40''$~[J2000];
$L_X{=}(3.7{\pm}0.8){\times}10^{44}$\,erg/s [0.1--2.4\,keV] -- Ebeling et
al.\ 1998).  These data reveal a tangential arc $2''$ from the optical
centroid of the BCG.  Spectroscopic observations at the Keck
Observatory confirm that the arc is a gravitationally--lensed galaxy at
$z{=}0.451$ (\S2).  The best--fit gravitational lens model faithfully
reproduces the arc morphology, however the total matter distribution of
this model appears to be much more elongated than the optical
isophotes of the BCG (\S3).  We also discuss the uniqueness of this
small--scale gravitational arc by comparing it with other known
cluster lenses, and outline how follow--up optical and X--ray
observations will help to fully exploit this powerful new constraint
on the matter distribution in cluster cores (\S4).  We assume
$H_0{=}65\,{\rm km~s^{{-}1}Mpc^{{-}1}}$, $\Omega_{\rm M}{=}0.3$ and
$\Omega_\Lambda{=}0.7$.  In this cosmology, $1''{\equiv}3.1{\rm kpc}$ at
$z{=}0.169$ and $1''{\equiv}6.2{\rm kpc}$ at $z{=}0.451$.
%\end{sloppypar}

\section{Observational Data and Analysis}

\subsection{\emph{Hubble Space Telescope} Imaging}

A\,1201 was observed through the F606W filter with \emph{HST} using
the WFPC2 camera on April 7, 2001.  We combined the $2{\times}400$--sec
exposures into a single mosaic using standard {\sc iraf} tasks, and
present in Fig.~1 the region of the
WF3 chip ($5''{\times}5''$) that contains the
central galaxy.  This frame reveals a tangentially distorted arc at a
radius of
$2''$ from the optical centroid of the BCG.  We interpret this arc as
arising from the gravitational distortion of a background galaxy by
the foreground cluster potential.

\begin{figure}
\centerline{
\psfig{file=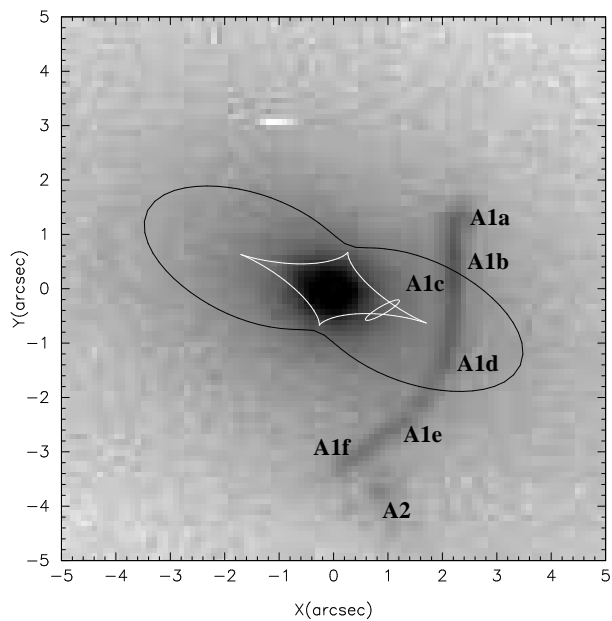,width=80mm,angle=0} }
\caption{ 
%{\sc left} --- \emph{HST}/WFPC2 view of the central
%$75''{\times}75''$ of A\,1201, taken from the WF3 chip.  The
%newly discovered tangential arc lies $2''$ from the optical centroid
%of the BCG and would have been undetectable from the ground.  {\sc right} ---
Zoom into the central $5''{\times}5''$ of the WF3 chip,
showing the detailed morphology of the arc and the multiple--image
identification upon which our lens model is
based (\S3).  The solid black curve shows the $z{=}0.451$ tangential critical
curve and the white astroid shows the $z{=}0.451$ caustic curve in the
source plane.  The white ellipse shows a schematic view of the
orientation and position of the galaxy in the source plane and reveals
how it lies across the caustic, giving rise to the distinctive
observed arc morphology.  }
\label{hst}
\end{figure}

\subsection{Keck--ESI Spectroscopy}

The arc was observed with the Echelle Imaging Spectrograph (ESI --
Sheinis et al.\ 2002) on the Keck--II 10--m telescope on the night of
April 12, 2002, in $0.6''$ seeing.  These observations and the
reduction of the data are described in detail by Sand et al.\ (2003).
In summary, a $1.25''{\times}20''$ slit was centered on the BCG,
oriented to intersect the portions of the arc labeled A1b and A1c in
Fig.~1.  The total integration time was 7.2--ksec.  The final reduced
spectrum covers the wavelength range $5300{\le}\lambda_{\rm
obs}{\le}9800$\AA, and contains [{\sc oii}~(3726,3729)],
H$\beta$, [{\sc oiii}~(4959,5007)] and H$\alpha$ at $\lambda_{\rm
obs}{=}(5406.7,5410.8)$, 7053.5, (7195.3,7264.7), 9523.1\AA
\,respectively, in addition to numerous other emission and absorption
features.  From these spectral features, we derive a redshift of
$z{=}0.451$ for the arc, thus confirming the gravitational lensing
interpretation.

\subsection{Multiple--image Interpretation}

We label the different segments of the arc (A1) in Fig.~1, together
with the faint feature that lies within $1''$ of the lower end of the
arc (A2).  The relatively high surface brightness of A1b/c and the dip
in surface brightness between this pair implies that they are a pair
of fold images arising from a portion of the galaxy that straddles the
$z{=}0.451$ caustic in the source--plane.  The de--magnified counter--image
of this pair would then most likely be A1f.  The position angle break
between A1d and A1e probably arises from a combination of the
source--plane morphology of the lensed galaxy and its orientation
relative to the caustic curve.  The emission from these two regions
has insufficient signal--to--noise to support definitive statements
regarding the multiple--image interpretation of A1d/e and indeed A1a.
However given their similarities in surface--brightness, it is
plausible that A1a/d/e are three images of the same portion of the
galaxy.  

\section{Gravitational Lens Modeling}\label{model}

\subsection{A simple ``relaxed'' model}

We construct a model of the projected mass distribution in A\,1201
using the {\sc lenstool} software developed by Kneib (1993; see also
Kneib et al.\ 1996).  The model consists of a single lens plane at
$z{=}0.169$, comprising two mass components (cluster--scale dark
matter halo and BCG) that we parametrize as truncated
pseudo--isothermal elliptical mass distributions (PIEMD -- Kassiola \&
Kovner 1993).  The projected cluster mass distribution is therefore
described by fourteen parameters: $x_{\rm c}, y_{\rm c}, \epsilon,
\theta, r_{\rm core}, r_{\rm cut}, \sigma_0$ for each of the two mass
components.  We match the central coordinates $(x_{\rm c}, y_{\rm c})$
for each component to the optical centroid of the BCG, as measured
from the \emph{HST} frame.  We also match the ellipticity
($\epsilon{=}0.23{\pm}0.03$) and position angle
($\theta{=}{-}(21{\pm}1)^\circ$) of these two mass components to that
of the observed light distribution measured at $R{=}2''$.  Typical
values of $r_{\rm core}$ and $r_{\rm cut}$ for cluster--scale mass
components are 50--$100{\rm kpc}$ and ${\gs}500{\rm kpc}$ respectively
(Smith 2002) -- i.e.\ well beyond the physical scales probed by the
arc in this cluster.  We therefore fix these two parameters at $75{\rm
kpc}$ and $1000{\rm kpc}$ respectively -- we find that none of the
results described below are sensitive to these choices.  This leaves
just four free parameters: the central velocity dispersion
$(\sigma_0)$ of the cluster and the core radius, cut--off radius and
central velocity dispersion $(r_{\rm core}, r_{\rm cut}, \sigma_0)$ of
the BCG.

We first constrain these parameters using the image pair A1b/c and the
location of the $z{=}0.451$ critical line that bisects these images.
This model is an acceptable fit to these constraints ($\chi^2/{\rm
dof}{\simeq}1$ -- see Smith 2002 for a detailed explanation of how
goodness--of--fit is estimated for such lens models), however it
predicts that the counter--image of A1b/c lies ${\sim}0.5''$ closer to
the center of the lens than the observed location of the candidate
counter images (i.e.\ A1e and A1f).  When A1f is added to the model
constraints, the fit deteriorates significantly ($\chi^2/{\rm
dof}{\simeq}100$).  The most straight--forward way to improve the
quality of this fit is to make the mass distribution more elliptical.
We therefore include the ellipticity of the cluster--scale dark matter
halo as a free parameter in the fit.  This yields an acceptable fit
for values of $\epsilon_{\rm DM}{\ge}0.7$.  This lower limit on the
cluster ellipticity is insensitive to whether A1e or A1f are adopted
as the third counter image of A1b/c.  We also
experiment with holding the ellipticity of the dark matter fixed at
$\epsilon_{\rm DM}{=}0.23$, and fitting for the ellipticity of the
BCG, obtaining $\epsilon_{\rm BCG}{\ge}0.7$.  Fixing $\epsilon_{\rm
DM}{=}\epsilon_{\rm BCG}$ and fitting for the ellipticity of the total
matter distribution, we obtain $\epsilon_{\rm DM}{=}\epsilon_{\rm
BCG}{\ge}0.5$.  In summary, the underlying cluster total mass
distribution appears to be significantly more elliptical than the
spatial distribution of stars in the BCG.

\begin{table}
\caption{\hfil Fiducial Lens Model Parameters\hfil }
\smallskip
\label{pars}
\begin{tabular}{lccccccc}
\hline\hline
\noalign{\smallskip}
{Mass}     & $x_{\rm c}$ & $y_{\rm c}$ & $\epsilon^{(1)}$ & {$\theta^{(2)}$} & {$r_{\rm core}$} & {$r_{\rm cut}$} & {$\sigma_o$} \cr
{Component}& (arcsec) & (arcsec) &      & {(deg)}    & {(kpc)}          & {(kpc)}         & {(${\rm km/s}$)} \cr
\noalign{\smallskip}
\hline
\noalign{\smallskip}
Cluster        & $0.0$ & $0.0$ & $>${\bf 0.7}$^{(3)}$ & ${-}21$ & $75$ & $1000$  &
{\bf 904}\cr
BCG            & $0.0$ & $0.0$ & $0.23$ & ${-}21$ & {\bf 0.7}  & {\bf
  150} & {\bf 197} \cr
\noalign{\smallskip}
\hline
\noalign{\smallskip}
\end{tabular}\\
{\footnotesize
$^1$ $\epsilon{=}(a^2{-}b^2)/(a^2{+}b^2)$ where $a$ and $b$ are the semi--major
  and semi--minor axes respectively.\\
$^2$ $\theta$ is measured anticlockwise from the positive $X$--axis in
  Fig.~1.\\
$^3$ Quantities in bold are free parameters in the lens
  model.
}
\end{table}

\subsection{Is A\,1201 Bi-modal?}

We also explore the possibility that A\,1201 is bi--modal, and examine
the \emph{HST} data for evidence of a second cluster--scale mass
clump.  An ${\sim} L^\star$ cluster galaxy lies on the WF2 chip in the
same direction as the position angle of the cluster mass distribution,
suggesting that a second mass clump may be associated with this
galaxy.  However, there are no other bright cluster members in this
vicinity, indicating that this scenario is quite unlikely (Smith et
al.\ 2002).  Weak shear maps may also be used to infer the likely
morphology of cluster mass distributions (e.g.,\ Kneib et al.\ 1996),
however the short exposure time of these \emph{HST} data preclude such
an analysis for A\,1201 (we estimate that just ${\sim}100$ suitable
faint galaxies are available across the entire WFPC2 field of view).
Despite the weak evidence for its existence, we quantify how massive a
second cluster--scale dark matter halo would have to be in order to
explain the observed multiple--images.  We fix the ellipticity of the
central dark matter halo and BCG at $\epsilon{=}0.23$ and add a
circular dark matter halo at the position of the bright cluster galaxy
noted above.  The best--fit velocity dispersion of this dark matter
halo is $\sigma_o{=}1000{\rm km/s}$ (with $r_{\rm cut}$ and $r_{\rm
core}$ held fixed at $1{\rm Mpc}$ and $50{\rm kpc}$ respectively).  In
this bimodal model, the velocity dispersion of the central dark matter
halo is ${\sim}750{\rm km/s}$, suggesting that if A\,1201 is bimodal,
then the dominant mass component would not be coincident with the BCG.

In summary, although a bi--modal mass distribution is allowed by the
current shallow \emph{HST} data, we suggest that an elliptical mass
distribution is the more likely explanation of the strong lensing
signal in this cluster.  We also note that our forthcoming
\emph{Chandra} observations (PID: 04800980, PI: Edge) will be the
first pointed X--ray observations of this cluster.  The X--ray
pass--band therefore currently offers no clues on the cluster mass
distribution.

\subsection{The Fiducial Model}

We adopt the model described in \S3.1 in which $\epsilon_{\rm
BCG}{=}0.23$ and $\epsilon_{\rm DM}{\ge}0.7$ as the fiducial lens
model and list the relevant parameters in Table~1.  We ray--trace each
portion of the arc through the fiducial model back to the
source--plane to double check our interpretation of the
multiple--images.  We summarize this exercise with the white ellipse
in Fig.~1 which shows the position, size and orientation of the galaxy
in the source--plane.  A1 therefore appears to be an elongated galaxy,
possibly an edge--on star--forming disk galaxy.  The observed
morphology of the arc may therefore be explained by a combination of
the elongated source--plane morphology and the orientation of this
galaxy relative to the caustic which we over--plot as the white
astroid in Fig.~1.  Integral field unit spectroscopic observations of
this arc (e.g.,\ Swinbank et al.\ 2003) would help to confirm our
interpretation of the arc as arising from an edge--on galaxy.

We also use the fiducial model to measure the projected mass enclosed
within the $z{=}0.451$ tangential critical curve, obtaining
$M(R{\le}2''){=}(5.9^{{+}0.9}_{{-}0.7}){\times}10^{11}{\rm M_\odot}$,
where the uncertainty is estimated from a family of lens models that
satisfy $\Delta\chi^2{\le}1$.  We identify these models by exploring the
five--dimensional parameter space defined by the free parameters in the
fiducial best--fit model.  The observed magnitude of the BCG in the same
aperture is $V_{606}(R{\le}2''){=}17.5{\pm}0.1$.  Correcting to the
observed $V$--band and applying both $k$--correction and galactic
extinction (Sand et al.\ 2003) we obtain
$M_V(R{\le}2''){=}{-}22.2{\pm}0.2$.  The total mass--to--light ratio of
A\,1201 on the scales probed by the tangential arc projected along the
line of sight is therefore $M_{\rm tot}/L_V{=}9.4^{{+}2.4}_{{-}2.1}({\rm
M/L})_\odot$.  This number is larger than values typical of stellar
populations of early--type galaxies (e.g.,\ Gerhard et al.\ 2001).  Indeed,
the joint lensing and dynamical analysis of this cluster by Sand et al.\
(2003) yields a stellar mass--to--light ratio of
$M_{\star}/L_V{=}3.8{\pm}0.3({\rm M/L})_\odot$.  We therefore conclude
that $60\%$ of the mass within the cylinder of radius $2''$, i.e. $15\%$
of the effective radius, is in the form of dark matter (i.e. the
ratio of stellar to total mass to light ratios).

%We also use the fiducial model to measure the projected mass enclosed
%within the $z{=}0.451$ tangential critical curve, obtaining
%$M(R{\le}2''){=}(5.9^{{+}0.9}_{{-}0.7}){\times}10^{11}{\rm M_\odot}$,
%where the uncertainty is estimated from a family of lens models that
%satisfy $\Delta\chi^2{\le}1$.  We identify these models by exploring
%the five--dimensional parameter space defined by the free parameters
%in the fiducial best--fit model.  The observed magnitude of the BCG in
%the same aperture is $V_{606}(R{\le}2''){=}17.5{\pm}0.1$.  Correcting
%to the observed $V$--band and applying both $k$--correction and
%galactic extinction (Sand et al.\ 2003) we obtain
%$M_V(R{\le}2''){=}{-}22.2{\pm}0.2$.  The mass--to--light ratio of
%A\,1201 on the scales probed by the tangential arc projected along the
%line of sight is therefore $M/L_V{=}9.4^{{+}2.4}_{{-}2.1}({\rm
%M/L})_\odot$.  This number is larger than values typical of stellar
%populations of early--type galaxies (e.g.,\ Gerhard et al.\
%2001). Indeed, the joint lensing and dynamical analysis of this
%cluster by Sand et al.\ (2003) yields a stellar mass--to--light ratio
%of $M/L_V{=}3.8{\pm}0.3({\rm M/L})_\odot$.  Adopting this value, we
%find that 60\% of the mass within the cylinder of radius $2''$,
%i.e. $15\%$ of the effective radius, is in the form of dark matter.

\section{Discussion and Conclusions}

\subsection{Is A\,1201 Unique Among Cluster Lenses?}

A\,1201 is drawn from a sample of $55$ X--ray luminous clusters
observed with \emph{HST} in the Edge et al.\ (PIDs: 8301 \& 8719)
snapshot survey of BCGs.  This snapshot survey doubles the number of
clusters that have been observed to date with either the WFPC2 or ACS
cameras.  Among these ${\sim}100$ clusters, A\,1201 is the only system
with a tangential arc on scales as small as $R{=}2''$.  This rarity
underlines the importance and efficiency of snapshot surveys with
\emph{HST} to discover such small--scale probes of the mass
distribution in clusters.

\begin{figure}
\medskip
\centerline{
\psfig{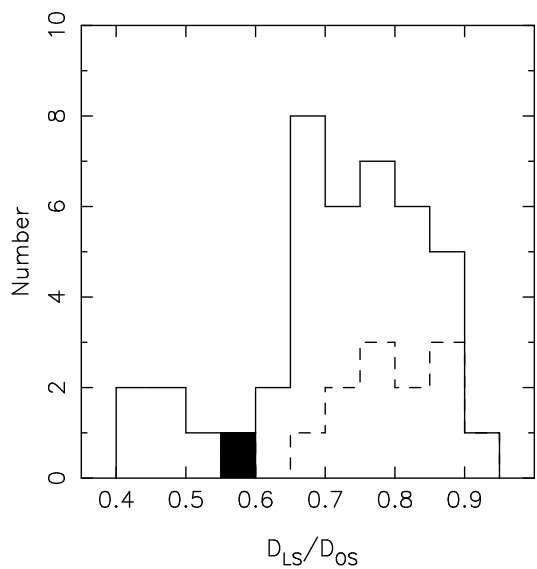}
}
\caption{The distribution of $D_{\rm LS}/D_{\rm OS}$ for observed
  multiple--image systems, drawn from an extensive search of the
  \emph{HST} archive and the published literature (Sand et al.\ 2004,
  in prep.).  The solid histogram shows the full sample of 41
  multiple--image systems; the dashed histogram shows a sample drawn
  from a well--defined sample of X--ray luminous clusters (Smith
  2002); the filled histogram marks the location of A\,1201.}
\label{hist}
\end{figure}

We also investigate the uniqueness of the tangential arc in A\,1201
among multiple--image systems in spectroscopically confirmed cluster
lenses.  The deflection angle of a  gravitational lens depends on 
the angular diameter distance ratio $D_{\rm LS}/D_{\rm OS}$
where $D_{\rm LS}$ is the distance from the lens to the source and
$D_{\rm OS}$ is the distance from the observer to the source.  We plot
the distribution of distance ratios for known multiple--image systems
in Fig.~\ref{hist}, based on an extensive review of the \emph{HST}
archive and the published literature (see Sand et al.\ 2004, in prep.\
for more details).  The cluster sample upon which this histogram is
based is heterogeneous.  We therefore also plot (as the dashed
histogram) the distribution of distance ratios for the multiple--image
systems identified thus far in a well--defined sample of X--ray
luminous clusters at $z{=}0.21{\pm}0.04$ by Smith (2002; see also Smith et
al.\ 2001, 2002; Kneib et al.\ 2003, in prep.; Smith et al.\ 2003, in
prep.).  A\,1201 lies at the lower envelope of both multiple--image
samples, with a value of $D_{\rm LS}/D_{\rm OS}{=}0.597$.  
From a lens geometry perspective, A\,1201 is therefore unusual but
not unique among the multiple--image systems discovered to date.
Other low $D_{\rm LS}/D_{\rm OS}$ systems include famous
lensing clusters, for example the giant arc ($z{=}0.724$; Soucail et
al.\ 1988) in A\,370 ($z{=}0.370$).  However, the higher redshift of
this, and other clusters with low distance ratios renders the observed
deflection angle (${\sim}10$--$20''$) and the physical scales probed
(${\sim}60$--120\,kpc) much larger than that relevant to A\,1201
($R{=}6$kpc).  A\,1201 is therefore unique in the small physical scales
probed by its tangential arc.

\subsection{Summary and Outlook}

\emph{HST} snapshot imaging of A\,1201 ($z{=}0.169$) with the WFPC2
camera reveals a tangential arc 2--arcsec from the center of this
cluster.  Spectroscopy obtained with ESI on the Keck--II telescope
confirms the gravitational nature of the arc, and places the source
galaxy at $z{=}0.451$.  We construct a gravitational lens model that
is able to reproduce the observed arc morphology.  The key feature of
this model is that the total matter distribution is significantly more
elongated ($\epsilon_{\rm total}{\ge}0.5$) than the light distribution
of the BCG on $2''$ scales ($\epsilon_{\rm BCG}{=}0.23{\pm}0.03$).
With the current data we are unable to determine whether the matter
distribution really is more elongated than the stellar distribution,
or if there is a significant amount of mass whose center of mass does
not coincide with the BCG. This could indicate that the cluster is
dynamically less mature than the optical data suggest.

The proximity of the arc to the center of this cluster is unique among
cluster lenses, and provides an important constraint on the mass of
the cluster on very small physical scales.  We measure the projected
mass within the tangential arc to be
$M(R{\le}2''){=}(5.9^{{+}0.9}_{{-}0.7}){\times}10^{11}{\rm M_\odot}$,
and the $V$--band mass--to--light ratio to be
$M/L(R{\le}2''){=}9.4^{{+}2.4}_{{-}2.1}({\rm M/L})_\odot$, the angular
scale of $2''$ corresponding to a physical scale of 6\,kpc.  This
constraint, in conjunction with complementary high--resolution
space--based data from \emph{Chandra} and multi--color follow--up with
\emph{HST}/ACS will lead to substantial progress in understanding the
distribution of mass in this cluster.  Extending the unique
small--scale (6\,kpc) mass constraint out to larger scales
(${\sim}50$--500\,kpc) also promises an important role for A\,1201 in
the quest to understand the physical processes at play in galaxy
cluster cores.

\section*{Acknowledgments}

GPS thanks Jean--Paul Kneib for sharing his {\sc lenstool}
ray--tracing code and Chuck Keeton for discussions about lens
statistics.  We also thank Richard Ellis for assistance with the Keck
observations.  We are grateful for financial support from the Royal
Society (ACE, SWA) and NASA (DJS, TT) through grant HST--AR--09527.
TT acknowledges support from NASA through Hubble Fellowship grant
HF-01167.01.

\end{document}